# Preferred Synthesis of Armchair Transition Metal Dichalcogenide Nanotubes

Abid[1†], Luneng Zhao[2†], Ju Huang[3†], Yongjia Zheng[1†], Yuta Sato[4,5], Tianyu Wang[1], Dmitry Levshov[6], Lingfeng Wang[1], Qingyun Lin[7], Zhen Han[8], Chunxia Yang[1], Bill Herve Nduwarugira[1], Yicheng Ma[1], Yige Zheng[1], Hang Wang[1], Salman Ullah[1], Afzal Khan[1], Qi Zhang[9], Wenbin Li[3*], Junfeng Gao[2*], Bingfeng Ju[1], Feng Ding[10], Yan Li[8], Wouter Herrebout[6], Kazu Suenaga[5*], Shigeo Maruyama[11], Huayong Yang[1], Rong Xiang[1*]

[1]*State Key Laboratory of Fluid Power and Mechatronic Systems, School of Mechanical Engineering, Zhejiang University, Hangzhou 310027, China*

[2]*Key Laboratory of Materials Modification by Laser, Ion and Electron Beams, Ministry of Education, Dalian University of Technology, Dalian 116024, China*

[3]*Department of Materials Science and Engineering & Key Laboratory of 3D Micro/Nano Fabrication and Characterization of Zhejiang Province, Westlake University, Hangzhou 310030, China*

[4]*Research Institute of Core Technology for Materials Innovation, National Institute of Advanced Industrial Science and Technology (AIST), Tsukuba 305-8565, Japan*

[5]*SANKEN (The Institute of Scientific and Industrial Research), Osaka University, 8-1 Mihogaoka, Ibaraki, Osaka 567-0047, Japan*

[6]*Theory and Spectroscopy of Molecules and Materials, Department of Physics and Department of Chemistry, University of Antwerp, Antwerp 2610, Belgium*

[7]*Center of Electron Microscopy, State Key Laboratory of Silicon and Advanced Semiconductor Materials, School of Materials Science and Engineering, Zhejiang University, Hangzhou 310027, China*

[8]*College of Chemistry and Molecular Engineering, Peking University, Beijing 100871, China*

[9]*Center for Advanced Optoelectronic Materials, College of Materials and Environmental Engineering, Hangzhou Dianzi University, Hangzhou 310018, China*

[10] *Suzhou Laboratory, Suzhou, 215123, China*

[11]*Department of Mechanical Engineering, The University of Tokyo, Tokyo 113-8656, Japan*

[†]*These authors contributed equally*

* Corresponding authors: liwenbin@westlake.edu.cn; gaojf@dlut.edu.cn; suenaga-kazu@sanken.osaka-u.ac.jp; xiangrong@zju.edu.cn

In this work, we present the synthesis of transition-metal dichalcogenide (TMDC) nanotubes with a preferred chiral angle. $SnS_2$, $MoS_2$, and $WS_2$ are formed with high yield and structural purity inside the channels of boron nitride nanotubes. Atomic-resolution imaging, nano-area electron diffraction, and Circular Dichroism spectroscopy reveal that these synthesized TMDC nanotubes prefer to have an armchair configuration, with a probability up to 84%. Density functional theory reveals a negligible difference in the formation energy between armchair and zigzag nanotubes, suggesting that the chirality preference does not originate from the differences in structural stability. However, a detailed TEM investigation revealed that these TMDC nanotubes formed via a transition state of nanoribbons, and these nanoribbons are energetically more stable in a zigzag configuration. Subsequent machine learning potential molecular dynamics simulations verify that zigzag nanoribbons do roll up to form an armchair $SnS_2$ nanotubes. Finally, this "zigzag nanoribbon to armchair nanotube" transition process is directly observed in real time by *in-situ* transmission electron microscopy. This work demonstrates the first, but likely general, experimental strategy for synthesizing chirality-preferred TMDC nanotubes.

One-dimensional (1D) nanotubes exhibit extraordinary quantum phenomena such as 1D confinement and van Hove singularities, resulting in unique mechanical, optical, and electronic properties(*1–4*). Compared to the well-studied carbon nanotube (CNT), transition metal dichalcogenide nanotubes (TMDC NTs) offer a wide variety of possible material compositions, providing additional freedom for tuning various material properties from bandgap engineering to exciton–polariton interactions (*2*, *5–9*). In addition, unlike graphene, the TMDC lattice is less symmetric, giving rise to a stronger nonlinearity, which has been demonstrated recently in several 1D TMDC structures (*10–13*). However, controlling TMDC NTs' atomic structure, particularly their chirality (the "twist" of the rolled sheet), remains a fundamental challenge (*14*, *15*). It took more than two decades after the discovery of CNTs to achieve the selective growth of specific chiralities (*16–19*), while for TMDC NTs, the most successful demonstration was the synthesis of coherent stacking (i.e., unichiral-angle) in multi-walled $WS_2$ NTs (*15*, *20*, *21*). No reliable strategy for specific chirality has been proposed thus far. In this work, we synthesized $SnS_2$ NTs using a tailored CVD technique and achieved preferential synthesis of armchair-configuration at an enrichment up to 84%. This claim is supported by numerous sophisticated characterization methods, and the underlying formation mechanism is explained and verified by multiple computational approaches as well as *in situ* transmission electron microscopy. Further, we have successfully applied our synthesis strategy to $MoS_2$ and $WS_2$, which suggests armchair preference could be general for TMDC NTs.

To produce $SnS_2$ NTs, we developed a four-step synthesis process schematically illustrated in Fig. 1A. First, single-walled carbon nanotubes (SWCNTs) were utilized as the starting material and sacrificial template. Next, boron nitride nanotubes (BNNT) were formed on the outer surface of the SWCNTs (*22*, *23*). Subsequent oxidization in air leaves behind pristine BNNTs with inner diameters of 1-12 nm. Finally, $SnS_2$ NTs were grown within the BNNT inner channels (see Fig. S1 and **SI methods section** for more experimental details). In this growth process, the diameter of the inner channel is pre-determined by the initial SWCNTs, and plays a key role in the successful formation of $SnS_2$ NTs. Neither too small nor too large for producing nanotubes (details shown later). This is the reason why other BNNTs (e.g., commercial products), which typically have a smaller inner diameter, do not result in successful growth (Fig. S2).

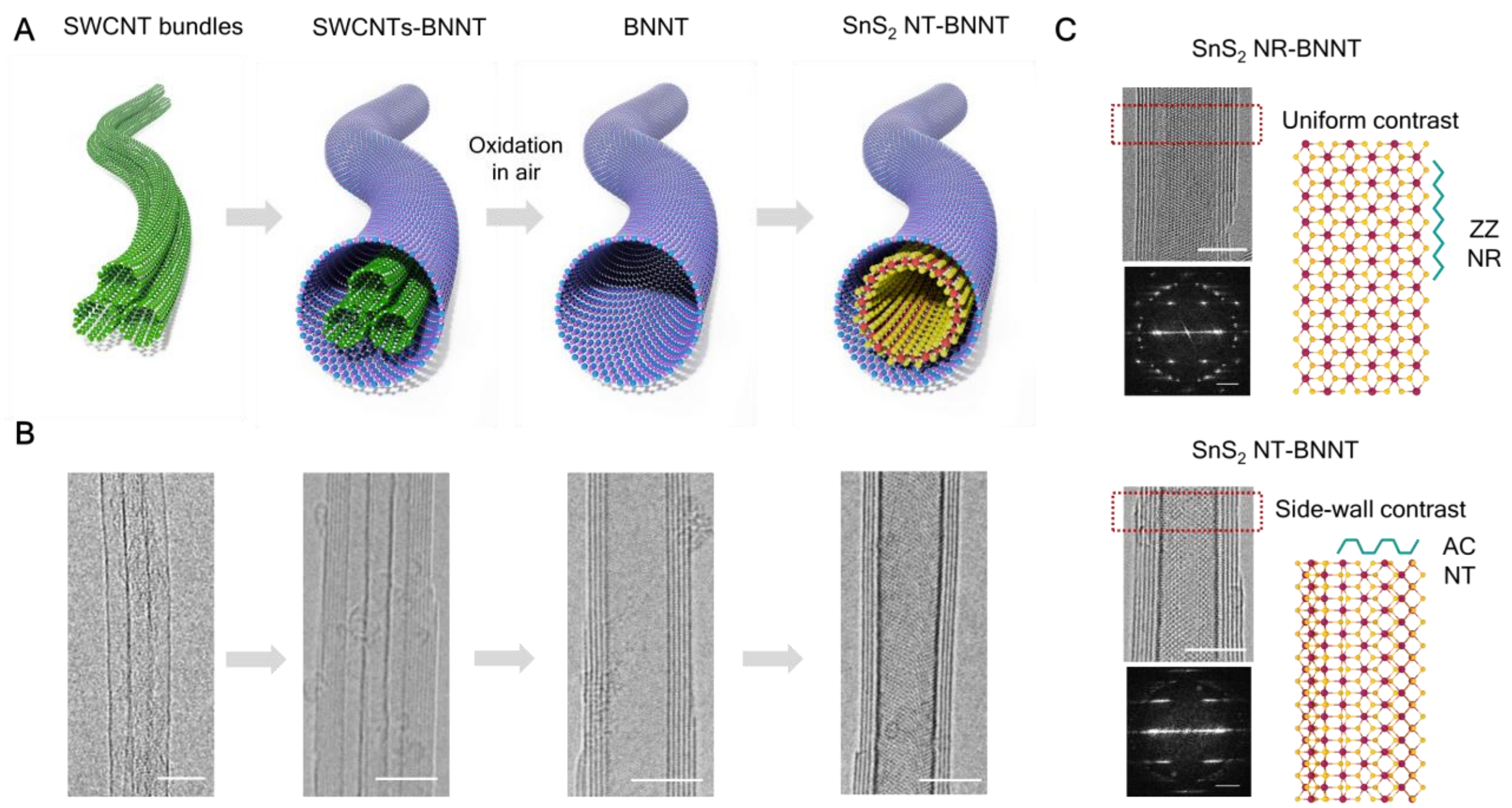

***Fig. 1. Synthesis process and structure of $SnS_2$ NR/NT within BNNT.*** *(**A**) An atomic model. (**B**) HR-TEM images present the stepwise synthesis of a $SnS_2$ NT within a BNNT. (**C**) HR-TEM images demonstrate the successful synthesis of $SnS_2$ nanostructures (NR and NT) within BNNTs, and the FFT shows that the NR is zigzag while the NT is armchair. All scale bars are 5 nm.*

Figure 1B shows high-resolution transmission electron microscopy (HR-TEM) images of the samples at different growth stages, with BNNT-encapsulated single-walled $SnS_2$ NTs as a final product. In representative TEM images (Figs. 1B and 1C), $SnS_2$ NTs inside the BNNTs show a strong contrast at side walls, which is a distinguishing feature of tubular crystals. Nanoribbons (NRs) are occasionally found inside the BNNTs, and can be differentiated from NTs because they generally exhibit uniform contrast along the entire plane (Fig. 1C (top)). The combination of these features allows one to easily distinguish NTs from NRs in TEM images. Moreover, in reciprocal space, a ribbon is a group of spaced dots, but these dots blend into prolonged dashed lines in the case of a NT due to the large curvature of the crystal basal plane. Figure 1C also shows the crystal structure and chiral definition of $SnS_2$ NRs and NTs. In NRs, chirality is conventionally defined following the atomic arrangement of the longer side, whereas for NTs, chirality is defined by the atomic arrangement of the edge perpendicular to the axis.

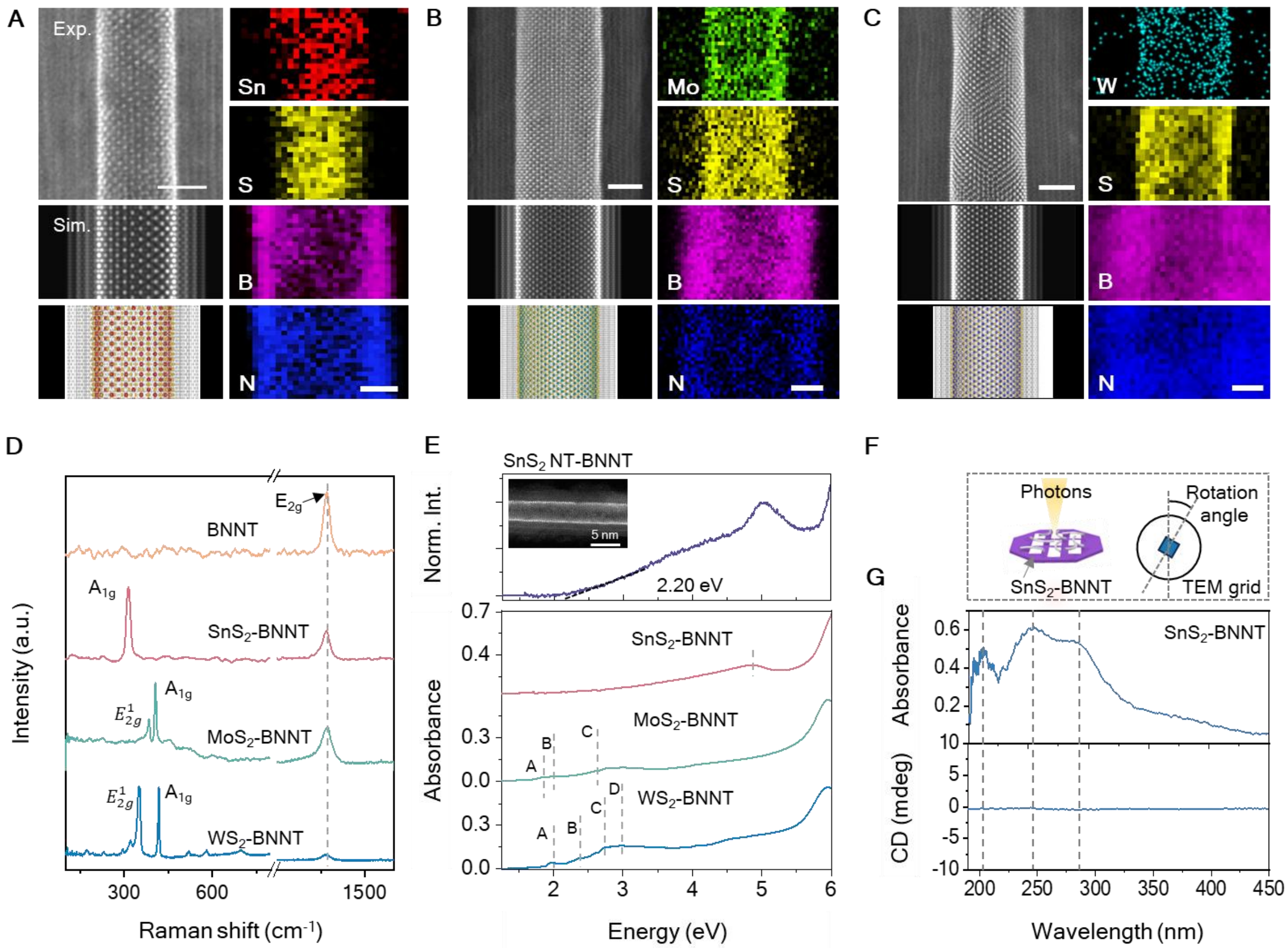

***Fig. 2. Comprehensive structural and spectroscopic analysis of $SnS_2$, $MoS_2$, and $WS_2$ encapsulated in BNNT.*** *HAADF-STEM images of individual* ***(A)*** *$SnS_2$,* ***(B)*** *$MoS_2$, and* ***(C)*** *$WS_2$ NTs inside a BNNT. Each image is accompanied by a simulated STEM image and an atomic model, confirming the successful synthesis and illustrating the atomic arrangement, overall morphology, and structural crystallinity. The right panel of each A, B, and C shows the corresponding EELS elemental maps, presenting the distribution of Sn/Mo, S, B, and N within the respective hybrid nanostructures. The EDS mapping was performed for W element due to the EELS limitations. All scale bars for the STEM images and for EELS mapping are 2 nm. (**D**) Raman spectra were collected for each material system: BNNT, $SnS_2$-BNNT, $MoS_2$-BNNT, and $WS_2$-BNNT, showing the characteristic vibrational modes of each TMDC, along with the BNNT mode.* ***(E)*** *Low-loss EELS spectra (top) for a very localized individual $SnS_2$ NT, and the UV-Vis absorption spectra (bottom) of $SnS_2$-BNNT, $MoS_2$-BNNT, and $WS_2$-BNNT, dashed lines indicate the characteristic features of each TMDC.* ***(F)*** *Schematic illustrating the geometry of the orientation-dependent CD measurements performed on the TEM grids.* ***(G)*** *Absorption (top) and electronic CD spectra (bottom) of the $SnS_2$-BNNT sample after subtracting the reference TEM grid signal.*

Figure 2 characterizes the structural, elemental, and optical properties of $SnS_2$ and other TMDC NTs using various spectroscopic and microscopic techniques. All results are obtained on as-grown samples without additional post-synthesis purification (Fig. S3). Figure 2A presents an experimental and simulated STEM-HAADF image, as well as EELS mapping of BNNT encapsulated $SnS_2$ NT. The shallow-focused STEM image reveals clear and periodic white dots, which correspond to Sn atoms in the upper surface of $SnS_2$ NT (*24*, *25*). The arrangement is in perfect agreement with 1T phase $SnS_2$ NT. Meanwhile, these Sn atoms are parallel to the tube axis, which only appears for armchair or zigzag NT. When a NT is chiral, Sn atoms will align with the tube axis at an angle between 0 to 30, and strong Moiré patterns are generated by the different contrasts of the top and bottom tube walls (Fig. S4). Elemental mapping using electron energy-loss spectroscopy (EELS) displays the spatial distribution of Sn, S, B, and N throughout the region. The whole tube contains Sn (red) and S (yellow) signals that are restricted to the BNNT inner channel. Additional STEM images and EDS/EELS mapping of the $SnS_2$ reveal that we obtained NTs with diameters ranging from 1.5 to 10 nm (Figs. S5-7). Similar results are obtained for $MoS_2$ and $WS_2$ NTs (Fig. 2B and 2C), in which heavy Mo and W atoms also nicely align with the tube axis, suggesting their achiral structures.

The chemical nature of the $SnS_2$ NTs was characterized by Raman scattering, optical, XPS, and EELS low-loss absorption spectroscopy (Figs. 2D and 2E, Figs. S6 and S8-S13), and the chirality of the $SnS_2$ NTs was evaluated by Circular Dichroism (CD) spectroscopy. Raman spectra of the sample at different stages, e.g., SWCNT, SWCNT-BNNT, BNNT, and $SnS_2$-BNNT, exhibit distinct features (Fig. S8A). BNNTs show a single $E_{2g}$ mode at 1369 $cm^{-1}$ (*22*, *26*), whereas in the final product, the $A_{1g}$ mode of $SnS_2$ at 314 $cm^{-1}$ becomes dominant (Fig. 2D) (*27*, *28*). The peak position also distinguishes 1T $SnS_2$ from possibly the 2H phase or other common compositions of Sn and S, e.g., SnS (*29*). Figure 2E compares the EELS low-loss absorption spectrum and the UV-Vis optical absorption spectrum. The former was obtained inside TEM on an individual $SnS_2$ NT, while the latter reflects the entire population of the film. Two different methods show high consistency. The overall chiroptical response of the film was further examined by circular dichroism (CD) spectroscopy. Figure 2F schematically illustrates the geometry of the orientation-dependent measurements on the TEM grids, and Figure 2G presents the absorption (top) and the corresponding CD (bottom) spectra of the $SnS_2$-BNNTs after subtraction of the grid background. The CD spectra were averaged over multiple orientations measured from both sides of the film to eliminate artifacts from linear dichroism or birefringence (*30*, *31*). Across the measured spectral range, the CD signal remains essentially zero (Fig. 2G, bottom), consistent with nano area electron diffraction (NAED) analysis (Fig. 3), which indicates that ~90% of the NTs are achiral (armchair + zigzag; see also Figs. S10 and S11).

Figure 3 demonstrates a collective evaluation of the chirality distribution of $SnS_2$ NTs. Previous studies have only achieved coherent stacking for multi-walled $WS_2$ NTs, but not demonstrated any preference for specific chirality for single-walled TMDC NTs. Electron diffraction is the most powerful tool and conclusively describes the atomic periodicity of a crystal by elastic scattering (*32–35*). Figure 3A displays a probability distribution graph that quantifies the statistical prevalence of different chirality types in $SnS_2$ NTs. Figure 3B presents the schematic illustration of the atomic structure of a monolayer 1T-$SnS_2$ and its respective NT formation by rolling at different helical angles. The schematic defines the chiral, zigzag, and armchair configurations relative to the rolling axis. Figure 3C presents the NAED pattern from a $SnS_2$ NT encapsulated in a BNNT, and its intensity profile is shown in Fig. S14. The diffraction pattern exhibits contributions from both materials: the $SnS_2$ reflections are highlighted with red hexagons, while the BNNT reflections are shown with a group of blue hexagons. We first note that all patterns are prolonged dashed lines rather than regularly spaced dots, confirming the curvature of the atomic sheets in $SnS_2$ NTs and BNNTs. Secondly, the alignment of the hexagons describes the chiral angle of the NTs. The apparently random orientation of the blue hexagons suggests that the outer BNNTs have no preferred chiral angle, however the singular inner red hexagon clearly reveals that the measured $SnS_2$ NTs exhibit a well-defined armchair chirality (discussion continues later). Similar patterns are obtained for $MoS_2$, and we performed simulations of NAED patterns to further validate the experimental observations. The simulated patterns of a single-walled $MoS_2$ NT demonstrate excellent agreement with the experimental ED pattern, thereby confirming the structural character. The diffraction patterns exhibit six-fold symmetry, characteristic of $SnS_2$ NTs, and corresponding to three distinct configurations: AC ($n = m$), ZZ ($n$ or $m = 0$), and chiral ($n \neq m$) (Fig. S15) (*36*). Besides the predominant AC NTs, Fig. 3(E and F) presents the atomic structure of $SnS_2$ NTs in these other two configurations, where hexagonal patterns align with the tube axis at different angles, clearly distinguishing the chirality of the measured NT.

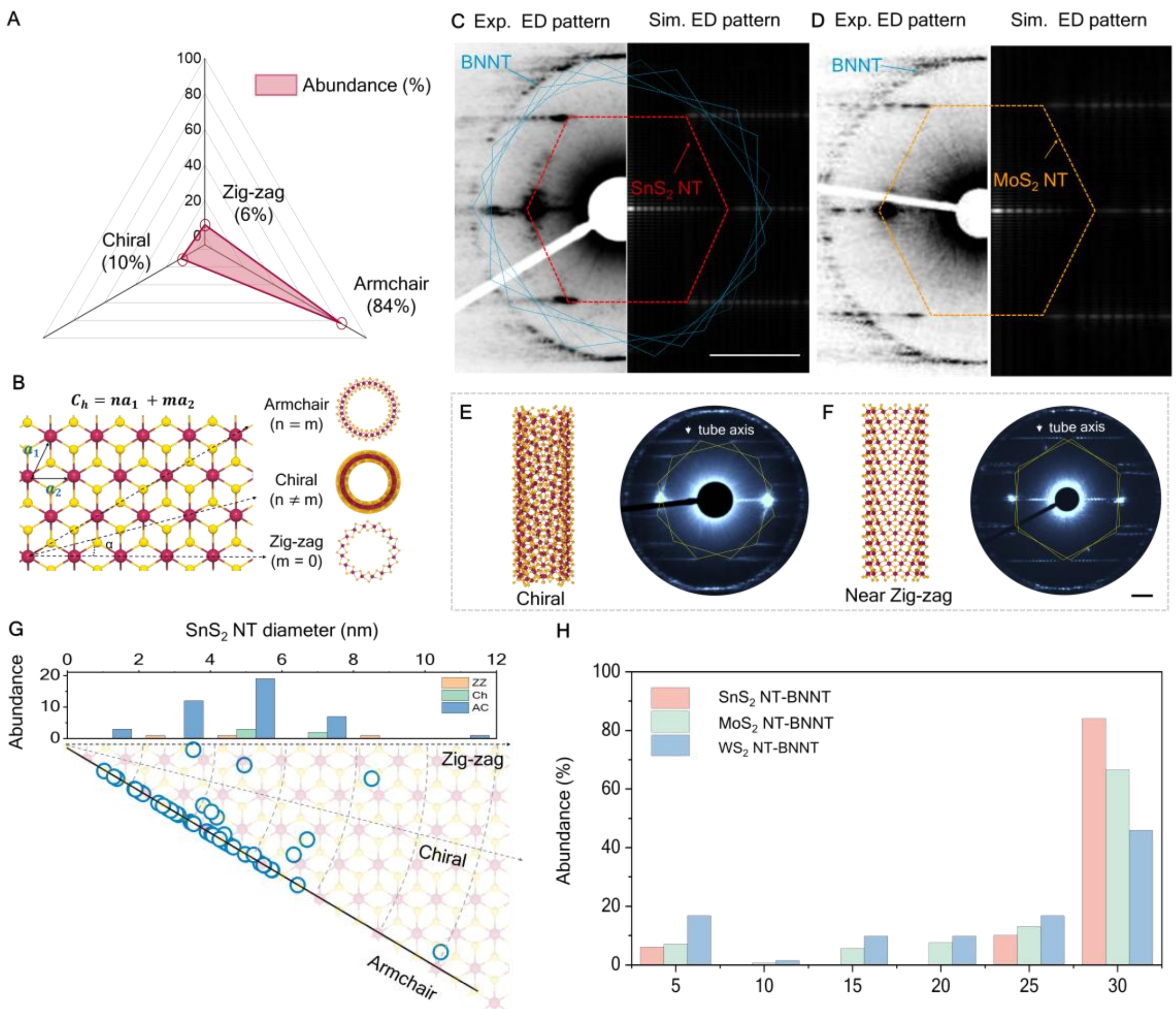

***Fig. 3. Chirality distribution for TMDC NTs (A)** A radar plot illustrates the distributions of various chirality types of $SnS_2$ NTs formed within the host nanotubes. **(B)** Schematic illustration of the atomic structure of a monolayer 1T-$SnS_2$ and its respective NT formation by rolling at different helical angles. **(C)** Experimental NAED patterns from $SnS_2$-BNNT. The red hexagon indicates the reflections from the $SnS_2$ NT, and the blue hexagons indicate the BNNT. **(D)** Experimental and simulated NAED patterns from $MoS_2$ NT-BNNT, with a hexagon indicating the reflections from the $MoS_2$ NT. **(E and F)** Atomic models of chiral and near-zigzag $SnS_2$ NTs, along with their corresponding NAED patterns. **(G)** Chiral angle (α) versus diameter for the statistically analyzed population of synthesized $SnS_2$ NTs. The schematic inset defines the chiral, zigzag, and armchair configurations relative to the tube axis. The majority of data points cluster near the armchair configuration (α ≈ 30°). **(H)** Statistical distribution of chiral angles for $SnS_2$ NTs, $MoS_2$ NTs, and $WS_2$ NTs. The bin size is 5°, where α < 5° corresponds to zigzag, 5° < α < 25° to chiral, and 25° ⩽ α ⩽ 30° to armchair configurations.*

We established statistical confidence by studying more than 50 NAED patterns; additional data presented in Fig. S16. Notably, out of 50 $SnS_2$ NTs, 42 exhibited AC structure, while five were chiral and three were near-ZZ. The analysis demonstrates a strong preference for $SnS_2$ NTs to adopt armchair chirality with an approximate probability of 84% among the nanotube products. For comparison, among the very few attempts found in literature, An et al. confirmed that different shells in multi-walled $WS_2$ NTs tend to have a uniform chiral angle within a single tube, and Nakanishi et al. found that the chirality of $MoS_2$ NTs is random (*14*, *37*). Meanwhile, all $SnS_2$ NRs coexisting here displayed ZZ configurations (Fig. S17, and reason to discuss later). Figure 3G presents the chiral angle (α) versus diameter for the statistically analyzed population of synthesized $SnS_2$ NTs. The BNNTs with 4-6 nm diameter lead to a higher yield of AC NTs. The majority of data points cluster near the armchair configuration ($\alpha = 30°$). For statistical analysis, we present the distribution of chiral angles for $SnS_2$, $MoS_2$, and $WS_2$ NT in Fig. 3H. The histogram is not uniformly distributed, and a large portion of the population of NTs is in the armchair range, indicating a strong preference for the armchair configuration in $SnS_2$ NT (84%) and $MoS_2$ NT (62.5%). However, $WS_2$ NT (45.8%) shows the armchair configuration, but to a lesser extent than $SnS_2$ and $MoS_2$ NT.

To understand the origin of the AC chirality preference of the three TMDC NTs, we investigated the stability of AC and ZZ configurations of NRs and NTs using DFT simulations, as shown in Fig. 4 and Fig. S19. In both 1T ($SnS_2$) and 2H ($MoS_2$ and $WS_2$) systems, the normalized energies are only dependent on the diameter, regardless of the chirality of NTs (Fig. 4B, and Fig. S19A and S19C). Therefore, the intrinsic NT stability alone cannot explain the experimentally observed chirality selection. In contrast, the edge energetics of TMDC NRs, (which are sometimes experimentally found attached to the end of the TMDC NTs (Figs. S29-30)), are independent of their width, but strongly depend on their chirality, i.e., the ZZ edge energies are much lower than those of AC edges for 1T $SnS_2$ (Fig. 4C). Besides the $SnS_2$, the ZZ edge also exhibits a clear and consistent lower formation energy than the AC configuration in both 2H $MoS_2$ (Fig. S19B) and $WS_2$ (Fig. S19D). Therefore, it is expected that the populations of ZZ NRs should be dominating in all explored systems ($SnS_2$, $MoS_2$, and $WS_2$) regardless of the 1T and 2H phase. If the AC NTs are cutting along the axial direction, they should be transformed into several ZZ edge NRs, and vice versa. It can be expected that connecting the ZZ edge to form the AC NTs, the dominating ZZ $SnS_2$ NRs will produce the dominant AC NTs, which is in agreement with experimental observations.

Nevertheless, the electron's effective mass ($m_e^*$) of $SnS_2$ AC NTs is noticeably lower than those of ZZ NTs across the entire explored diameter range, according to DFT calculations (Fig. 4E), implying the faster carrier mobilities in AC NTs in microelectronics. Besides, the electronic bandgap of both ZZ and AC NTs can be continuously tuned from 1.07 eV to 1.51 eV as the diameter increases from 1.7 nm to 6.1 nm (Fig. 4D and Fig. S20-21, S23-24, Table S4). While for the hole effective mass ($m_h^*$), 10 times heavier than $m_e^*$ in both AC and ZZ $SnS_2$ NTs (Fig. S20). When the diameter reaches 3 nm, the $m_h^*$ of AC NTs is also much lower than that of ZZ, implying the superiority of $SnS_2$ AC NTs over ZZ NTs. The similar electronic properties were also revealed in the $MoS_2$ and $WS_2$ systems (Figs. S20 and S22, and Figs. S25-28).

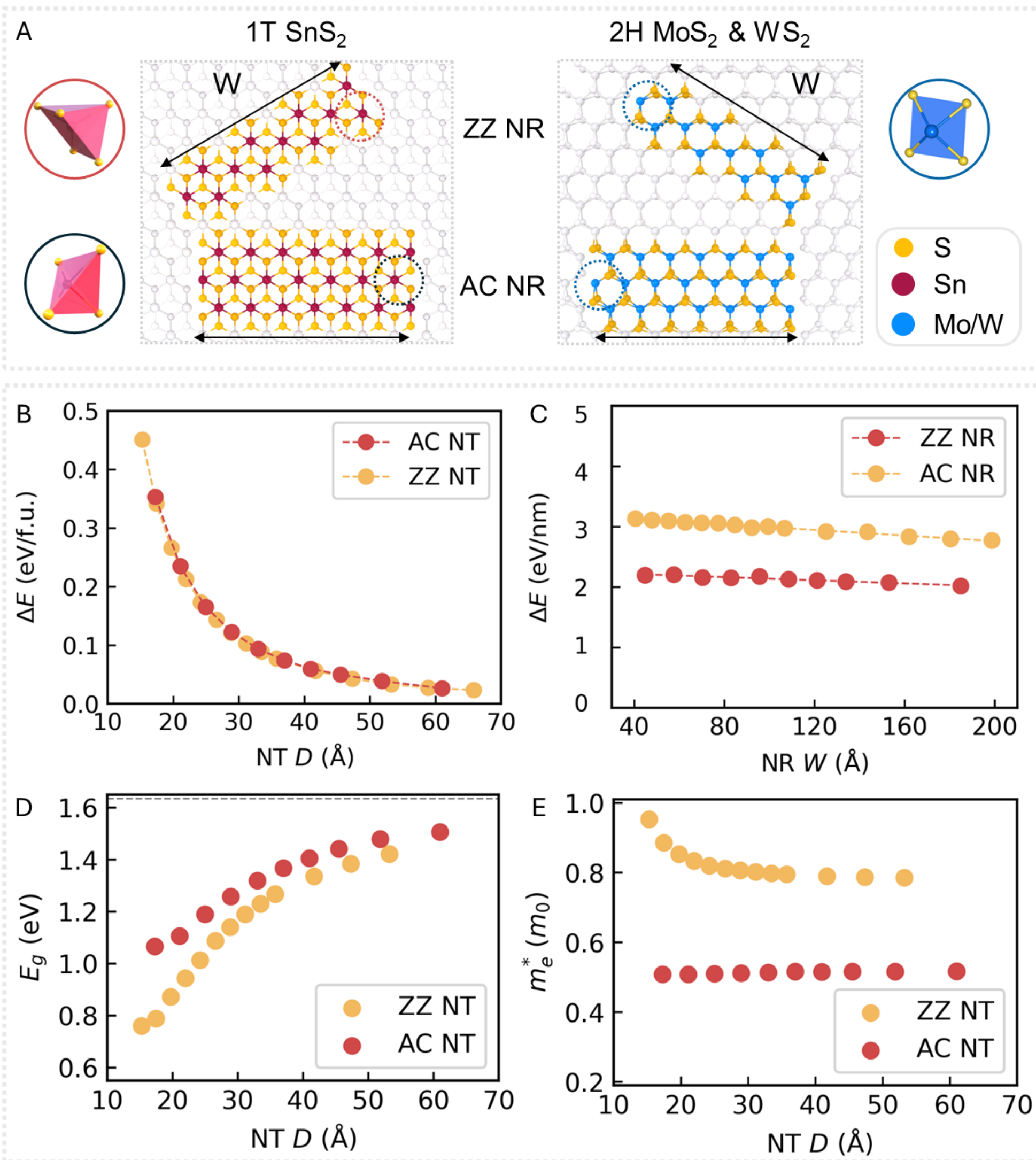

***Fig. 4. Energetic stability of TMDC NRs and NTs confined within BNNTs.*** ***(A)*** *Atomic structures of 1T-SnS2 and 2H-MoS2/WS2, along with schematic illustrations of their corresponding ZZ and AC NRs. The formation of ZZ and AC NRs from their monolayers is summarized in Table S1.* ***(B)***

*Normalized energy of NTs as a function of tube diameter for $SnS_2$. The NT energy was normalized to the energy of their respective monolayer (Table S2), and f.u. represents the formula unit of $SnS_2$* ***(C)*** *Normalized formation energies of NRs as a function of ribbon width (W) for $SnS_2$. Energies are normalized to the corresponding monolayer per unit length (1/L) along the periodic direction.* ***(D)*** *Electronic band gap of AC and ZZ $SnS_2$ NT, and* ***(E)*** *Effective mass of electron ($m_e^*$) for AC and ZZ $SnS_2$ NT***.**

To answer whether ZZ NRs can form the dominating $SnS_2$ AC NTs? An accurate simulation is critical for understanding the kinetics based on the possible evolution structures observed by HAADF and in-situ real-time TEM experiments. However, most traced structures, including over 100k atoms, are far beyond DFT capacity. Here, DFT data-driven machine learning potential (MLP), which possesses both large system and accuracy at the same time, is developed in our simulations. The details of the DFT dataset and training of the MLP are shown in Figs. S31 and S33A. The resulting MLP shows mean absolute errors (MAE) of energy and force of 12 meV/atom and 182 meV/Å, respectively (Fig. S33B-C), thus providing reliable results. In addition, representative structures extracted from the simulated transformation pathway ($SnS_2$ NR to NT) were further evaluated by DFT, and the resulting energy ordering agrees closely with that predicted by the MLP (Fig. S34), further confirming the reliability of the MLP in describing this dynamic process.

For the initial stage, a trilayer $SnS_2$ NR with BNNT template wrapping was simulated using MLP-MD (Fig. 5A; Fig. S32 and S35, and details in the Movie S1). The bending energy of TMDC is commonly much higher than that of flexible BN. $SnS_2$ is likely to form near-flat NRs adsorbed on the inner wall of the BNNT (Fig. S18), and introduce the wrapping wall of BNNT to be straight. Finally, the $SnS_2$ NRs will introduce the collapse of BNNT. This collapse process exhibits an energy barrier (~0.1 eV/Å) due to the bending rigidity of the BNNT. However, because the collapsed structure increases the vdW interaction between the BNNT and the $SnS_2$ NRs, total energy is reduced by 0.6 eV/Å (Fig. 5A). Once this collapsed BNNT-$SnS_2$ configuration forms, the number of layers in the $SnS_2$ NR is fixed, and the $SnS_2$ NR can only continue to grow along the axial direction of the BNNT template (as shown in Fig. 5B). In the whole process, the breathing-mode vibration of BNNT is frequency attach and detach $SnS_2$ NRs, which may exfoliate the $SnS_2$ NRs by the vdW force. When the ribbon is grown with a certain width and layers, the edge of $SnS_2$ NRs can connect to form $SnS_2$ NTs via exfoliating, interlayer sliding, and edge connecting processes (Fig. 5C and details in Fig. S34). This is similar to observations in CNT (*19*). But for flexible carbon NRs, twisting into spiral NRs and then connecting the edges to form chiral CNTs is more favorable (*19*), different from the transition from hard-bending $SnS_2$ NRs to NTs. It is also noted that the bending stiffness and layers of $SnS_2$ NRs are crucial to form uniform NTs without chirality. If the layers of $SnS_2$ NR are less than three, to match the BNNT diameter, chiral $SnS_2$ NTs will also form by similar twisting behaviors like CNTs. This also explains why a small number of chiral $SnS_2$ NTs are observed experimentally. A model including edge, van der Waals, and bending energies explains the observed $SnS_2$ NT diameter range (Fig. S36).

It is noted that the collapse of BNNT only occurs when it is located with inner $SnS_2$ NRs; the other part of BNNT is still cylindrical, which is like the collapse part of BNNT to recover the tubular structures. Such exfoliating, sliding, and edge connecting will occur on one side (green box in Fig 5D), and then propagate into another side. This process can be traced by a series of critical structures from MLP-MD simulations (Fig. 5D, Movies S2 and S3). The simulated STEM images of such critical structures (Fig. 5E) are compared with experimental STEM images (Fig. 5F). These images are in very good agreement with each other, strongly evidencing the aforementioned kinetic mechanism from $SnS_2$ ZZ NRs to AC NTs. This mechanism was further validated by real-time observation by in-situ TEM (Fig. 5G, Movie S4 for the entire process). HR-TEM snapshots captured the initial collapse of BNNT with inner $SnS_2$ NRs from 0 s, the breath-mode vibration of BNNTs, and recovery tubular configuration from the collapse in a later frame (45 s), which fully reproduces the simulated process shown in Fig. 5C and D.

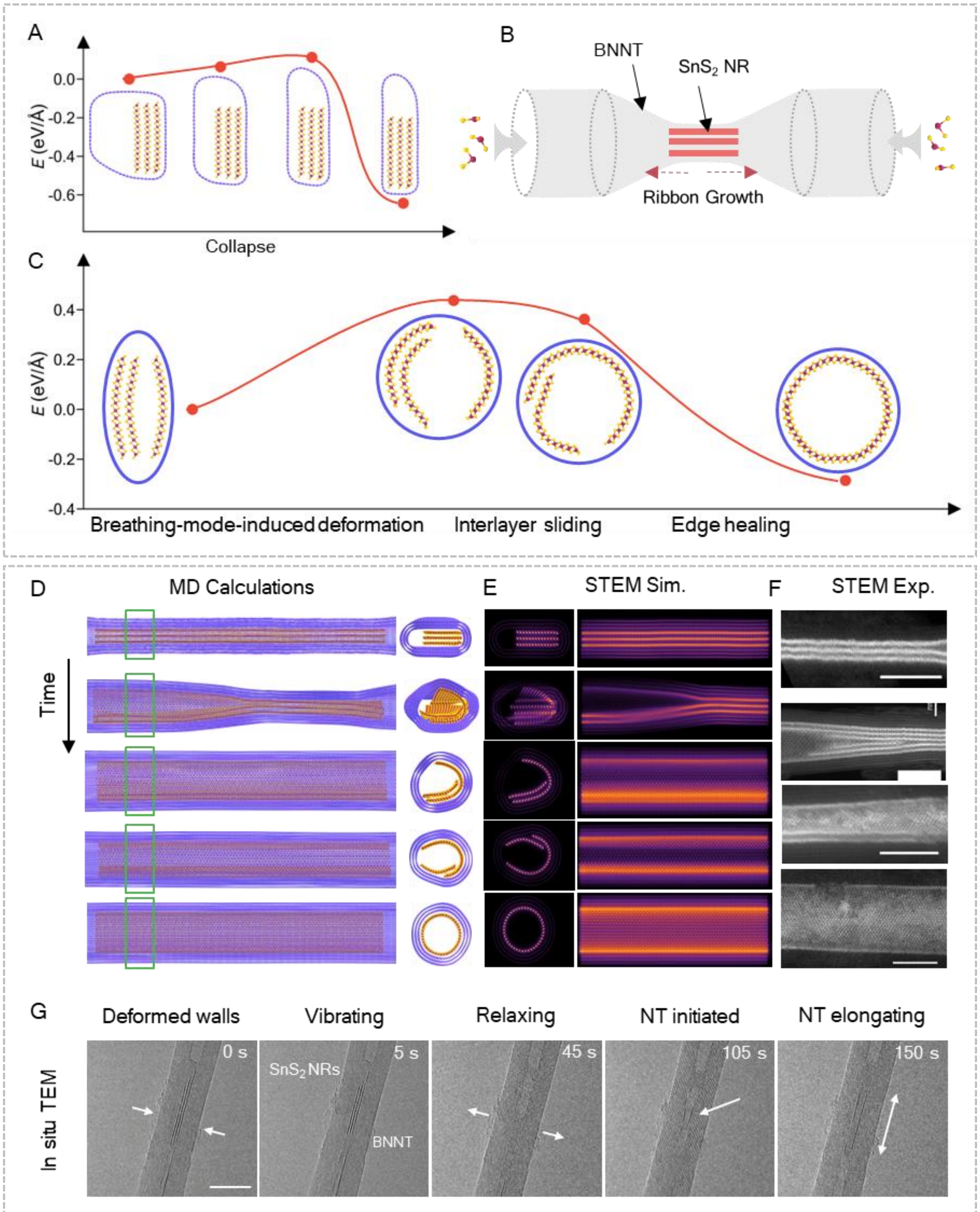

***Fig. 5. Confinement-induced deformation and nanoribbon-to-nanotube transformation of $SnS_2$ inside BNNTs.*** *(A) Progressive collapse of the BNNT template and the corresponding energy evolution for a trilayer $SnS_2$ nanoribbon confined inside the BNNT. (B) Schematic illustration of the deformation induced by interactions between the $SnS_2$ nanoribbons and the BNNT walls. (C) Proposed nanoribbon-to-nanotube transformation pathway of $SnS_2$, proceeding through*

*breathing-mode-induced deformation, interlayer sliding, and edge healing/closure to yield a closed nanotube. (D) Atomic models of representative intermediate structures along the transformation pathway. (E) Simulated STEM images corresponding to the models in (D). (F) Experimental STEM images showing closely matched intermediate morphologies, supporting the proposed mechanism. Scale bar, 5 nm. (G) Time-sequenced HR-TEM snapshots showing the transformation of* $SnS_2$ *nanoribbons into nanotubes under 200 kV electron-beam irradiation. All scale bars, 10 nm.*

We have demonstrated the preferred synthesis of armchair $SnS_2$ NTs within BNNT templates. Similar results are also obtained for $MoS_2$ and $WS_2$ NTs, suggesting the observed armchair preference could be general (Figs. S37-39, and details in movies S5 and S6). DFT, MLP-MD calculations, comparison of simulated and experimental HAADF-STEM images, and *in-situ* real-time HR-TEM observations have provided a consistent and robust explanation of the structural selectivity. This work provides us with a solution to the three-decades-pending issue for chirality control in TMDC nanotubes. The revealed mechanism may also inspire the controlled synthesis of other types of nanotubes. Armchair $SnS_2$ NTs have noticeably lower effective masses, thus potentially higher carrier mobility, and could be used for high-performance electronics.